\documentclass[prl,twocolumn,showpacs]{revtex4}
\usepackage{graphicx}
\usepackage{bm}
\usepackage{amssymb,amsfonts,amsmath}

\newcommand{\refeq}[1]{(\ref{#1})}
\newcommand{\reffig}[1]{FIG. \ref{#1}}

\newcommand{\ca}{$\mathrm{C}_\mathrm{\alpha}$}

\begin{document}
\title{Residue network in protein native structure belongs to\\
the universality class of three dimensional critical percolation cluster}
\author{Hidetoshi Morita}
\email{Hidetoshi.Morita@inln.cnrs.fr}
\altaffiliation[Present address: ]{
INLN, CNRS, 1361 route des Lucioles, 06560 Valbonne, France}
\author{Mitsunori Takano}
\email{mtkn@waseda.jp}
\affiliation{Faculty of Science and Engineering, Waseda University,
Tokyo 169-8555, Japan
}
\date{\today}

\begin{abstract}
A single protein molecule is regarded
as a contact network of amino-acid residues.
Some studies have indicated that this network is a small world network (SWN),
while other results have implied that this is a fractal network (FN).
However, SWN and FN are essentially different in the dependence of
the shortest path length on the number of nodes.
In this paper, we investigate this dependence
in the residue contact networks of proteins in native structures,
and show that the networks are not SWN but FN.
FN is generally characterized by several dimensions.
Among them, we focus on three dimensions;
the network topological dimension $D_c$,
the fractal dimension $D_f$,
and the spectral dimension $D_s$.
We find that proteins universally yield
$D_c \approx 1.9$, $D_f \approx 2.5$ and  $D_s \approx 1.3$.
These values are in surprisingly good coincidence with those
in three dimensional critical percolation cluster.
Hence the residue contact networks in the protein native structures
belong to the universality class of three dimensional percolation cluster.
The criticality is relevant to the ambivalent nature of the protein
native structures, i.e., the coexistence of stability and instability,
both of which are necessary for a protein to function
as a molecular machine or an allosteric enzyme.
\end{abstract}


\maketitle

\paragraph{Introduction}
Proteins are one-dimensional chains of amino-acid residues
embedded in three dimensional ($D=3$; 3D) Euclidean space.
The residues neighboring in the Euclidean space are
in contact with each other. Thus we can regard a protein molecule
as a contact network of amino-acid residues
\cite{Bahar_Atilgan_Erman1997,Atilgan_etal2001}.
This network viewpoint is complementary to the energy landscape picture
\cite{Frauenfelder_Sligar_Wolynes1991}
in understanding the general properties of proteins. 
We hereafter consider this network within single protein molecules
in their native structures,
in particular focusing on its universality among proteins.

Some recent studies~\cite{Vendruscolo_etal2002,Dokholyan_etal2002,
Greene_Higman2003,Atilgan_Akan_Canan2004,Bagler_Sinha2005}
have applied the latest network theory to the residue network,
by regarding the amino-acid residues and their contacts
as nodes and edges, respectively.
The important quantities to characterize the network are
the clustering coefficient $C$
and the shortest path length $L$~\cite{Newman2003_rev}.
Those studies have demonstrated that in the residue networks
$C$ is larger than the random networks~\cite{RN}
while $L$ is smaller than the normal lattice.
This indicates that the residue network
is a small world network (SWN)~\cite{Watts_Strogatz1998}.

On the other hand,
the spacial profile of residues within single protein molecules
has long been studied with the use of authentic methods of material science.
Earlier spectroscopic studies~\cite{spectr}
have shown anomalous density of states.
These results, accompanied with theoretical studies
\cite{spectr_theor}, have suggested
that the protein structures possess the property of fractal lattice.
The fractality within single proteins has also been supported numerically
through the density of normal modes
\cite{Wako1989,benAvraham1993,Yu_Leitner2003}
and the spacial mass distribution
\cite{frac_dim_prev}.
This implies that the residue network that we are interested in
is a fractal network (FN).

From the general viewpoint of the network theory, however,
there lies a dichotomy between SWN and FN~\cite{Csanyi_Szendroi2004}.
The clustering coefficient $C$ cannot discriminate between SWN and FN,
since in both the networks $C$ have a larger value than the random networks.
In contrast, the dependence of the shortest path length $L$
on the number of nodes $N$ is essentially different between SWN and FN;
$L$ depends on $N$ logarithmically and algebraically, respectively.
By exploiting the $N$-dependency of $L$,
we can differentiate SWN and FN, in principle.

In proteins, nevertheless, it is practically difficult to clearly
distinguish between these two $N$-dependence.
This is because the size of proteins does not distribute widely
enough to cover sufficient decades.
The same data sets can be read as a straight line
both in log-log (SWN) and semi-log (FN) plot.

To overcome this difficulty,
here we introduce a more sophisticated method.
Instead of the $N$-$L$ plot among various sized proteins,
we investigate an equivalent {\it within single protein molecules};
we calculate the number of nodes $n_l$
that can be reached until $l$ path steps.
Then, by overdrawing the $n_l$-$l$ plot for various sized proteins,
we obtain a universal curve,
as well as the deviation from it due to finite size effect.
Thus we can discuss an asymptotic behavior in the large $N$ limit.
We thereby find that network in protein native structures
is FN, not SWN. This is the first result of this letter.

We then obtain the three characteristic dimensions
of fractal residue network; the network topological dimension $D_c$,
the fractal dimension $D_f$, and the spectral dimension $D_s$.
The values of them are universal among single-chain proteins.
Furthermore, these three values surprisingly coincide
with those of the 3D critical percolation cluster.
Namely, proteins belongs to the universality class of
3D critical percolation cluster.
This is the second and the most highlighted result of this letter.

\paragraph{Small world network vs fractal network}
First of all, we define the network in a protein native structure.
We use the spacial information of the native structure
in Protein Data Bank (PDB)~\cite{PDB}.
We regard amino-acid residues as nodes;
we represent them by \ca~atoms, which is a standard way
in coarse grained models~\cite{Atilgan_etal2001},
and is indeed employed in the past network studies
\cite{Vendruscolo_etal2002,Dokholyan_etal2002,
Atilgan_Akan_Canan2004,Bagler_Sinha2005}.
A pair of nodes, $i$ and $j$, is considered to have an edge if
their Euclidean distance, $d_{ij}$, is less than a cut-off distance, $d_c$.
Then the network is characterized by the adjacent matrix:
\begin{align}
\mathbf{A}=(A_{ij}),
\quad
A_{ij}=\Theta(d_c-d_{ij})
\end{align}
where $\Theta(\cdot)$ is the Heaviside step function.
Here we adopt $d_c=$ 7{\AA}, which corresponds to
the second coordination shell in the radial distribution function of \ca;
we have also confirmed that the result below is robust
to the choice of $d_c$ from 6 to 10{\AA}~\cite{fullpaper}.

Let $n_l^{(i)}$ be the number of nodes that a walker on the network
starting from the node $i$ can visit at least once until $l$ steps.
Since we are interested in the overall network property of a protein,
we consider its average, $n_l=\sum_i n_l^{(i)}/N$.
As $l$ becomes larger, $n_l$ monotonically increases
and finally saturates at $N$.
In the $D$-dimensional normal lattice, $n_l \sim l^D$.
If the network is FN, similarly, the following scaling holds
\cite{Csanyi_Szendroi2004}:
\begin{align}
n_l \sim l^{D_c},
\label{eq:n_l_FN}
\end{align}
where $D_c$ is referred to as the network topological dimension
\cite{Stauffer_Aharony1994, Nakayama_Yakubo_Orbach1994}.
If the network is SWN, in contrast, the relationship is
\cite{Csanyi_Szendroi2004},
\begin{align}
n_l \sim \exp(l/l_0),
\label{eq:n_l_SWN}
\end{align}
for a positive constant $l_0$. Note again that the relationships
\refeq{eq:n_l_FN} and \refeq{eq:n_l_SWN} are essentially different,
leading to the dichotomy between FN and SWN~\cite{Csanyi_Szendroi2004}.

\begin{figure}[tb]
\includegraphics[width=0.48\textwidth]{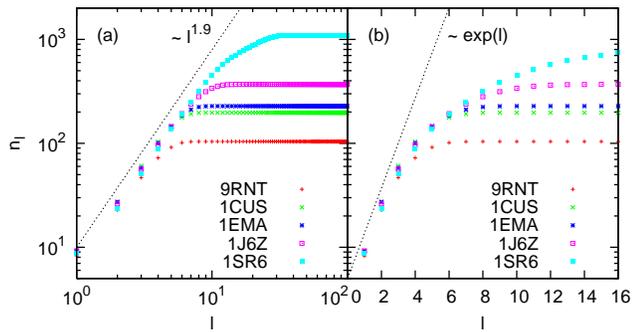}
\caption{Averaged number of nodes $n_l$ that a walker on the network
starting from a node can visit at least once until $l$ steps;
plotted in (a) log-log and (b) semi-log scales.}
\label{fig:n_vs_l}
\end{figure}

\reffig{fig:n_vs_l} shows the relationship between $n_l$ and $l$;
the same data sets are plotted in (a) log-log and (b) semi-log scales.
We present the data for five representative proteins of different size:
ribonuclease T1 (PDB ID=9RNT, 104 amino acids (a.a.)),
cutinase (1CUS, 200 a.a.),
green fluorescent protein (1EMA, 236 a.a.),
actin, (1J6Z, 375 a.a.),
and subfragment 1 of myosin (1SR6, 1152 a.a.).
Obviously the data obeys the power-law scaling
better than the exponential dependence.
This is also supported by considering the finite size effect as follows.
In (a), the range where the data follow the power-law scaling tends to extend
as the number of nodes $N$ increases.
This suggests the existence of an asymptotic universal line
in the limit of $N\to\infty$.
In (b), on the contrary, we cannot see such an asymptotic tendency.
Thus we conclude that proteins universally obeys the power-law scaling
\refeq{eq:n_l_FN} with $D_c\approx 1.9$.
Hence the networks in protein native structures are FN, not SWN.

In much larger proteins,
$D_c$ often gives a bit larger value than 1.9,
or even the scaling itself is smeared.
This is because the larger proteins are usually
not single-domain nor single-chain but multi-domain or multi-chain proteins.
Even in such proteins, however, each single-domain or single-chain component
still yields the same scaling law with the same dimension $D_c\approx 1.9$
\cite{fullpaper}.

One plausible reason why the network is not SWN but FN is that
the residues are spatially restricted in the 3D Euclidean space.
Indeed, it has been suggested that networks
with spatial (geographical) restriction tend to be rather regular
(including fractal) network than SWN~\cite{Csanyi_Szendroi2004}.

\paragraph{Fractal dimension}
In addition to the network topological dimension $D_c$,
FN is characterized by two other dimensions in general;
the fractal dimension $D_f$
and the spectral dimension $D_s$~\cite{Stauffer_Aharony1994}.
While these three dimensions and the Euclidean dimension $D$
are identical in the normal lattice, they can be different in FN.

The fractal dimension is determined from the spacial distribution of nodes.
Here we again employ the method within single proteins, differently from
the previous studies~\cite{frac_dim_prev},
in order to discuss the asymptotic behavior in the limit $N\to\infty$.
Let $n^{(i)}(d)$ be the number of nodes
the distance of which from the node $i$ is less than $d$;
$n^{(i)}(d)=\sum_j \Theta (d-d_{ij})$.
Since we are interested in the overall network property of a protein,
we consider its average, $n(d)=\sum_i n^{(i)}(d)/N$, that is,
\begin{align}
n(d)=\frac{1}{N}\sum_{i=1}^{N}\sum_{j=1}^{N} \Theta (d-d_{ij}).
\end{align}
Note that this is nothing but the correlation integral
introduced by Grassberger and Procaccia~\cite{Grassberger_Procaccia1983a},
although this is not normalized in order to consider the finite size effect.
As $d$ becomes larger, $n(d)$ monotonically increases
and finally saturates at $N$.
In the $D$-dimensional normal lattice, $n(d) \sim d^D$.
Similarly, if the spacial distribution of nodes is fractal,
\begin{align}
n(d) \sim d^{D_f},
\label{eq:n_vs_d_FN}
\end{align}
where $D_f$ is referred to as the fractal dimension.

\begin{figure}[tb]
\begin{center}
\includegraphics[width=0.35\textwidth]{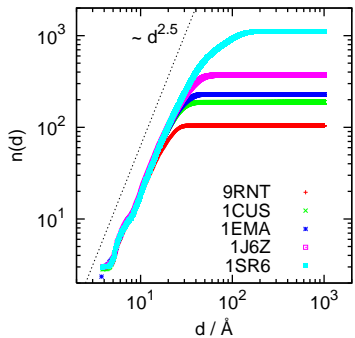}
\caption{Averaged number of residues $n(d)$
the distance of which from a residue is less than $d$
for the same proteins as
\reffig{fig:n_vs_l}.}
\label{fig:n_vs_d}
\end{center}
\end{figure}

\reffig{fig:n_vs_d}
shows $n(d)$ versus $d$ in log-log scale,
for the same proteins as \reffig{fig:n_vs_l}.
The relationship follows power-law scaling.
Similarly to the case in the network topological dimension,
this is supported by considering the finite size effect;
the power-law scaling range tends to extend as the number of nodes increases,
suggesting the existence of an asymptotic universal line
in the limit $N\to\infty$.
Thus we conclude that proteins universally follows the power-law scaling
\refeq{eq:n_vs_d_FN} with the fractal dimension $D_f\approx 2.5$,
which is consistent with the previous studies~\cite{frac_dim_prev}.

\paragraph{Spectral dimension}
The spectral dimension is determined
from the density of normal modes (DNM).
According to the Debye theory,
DNM in $D$-dimensional normal lattice is $\rho(\omega) \sim \omega^{D-1}$.
Similarly, DNM in FN obeys,
\begin{align}
\rho(\omega) \sim \omega^{D_s-1},
\label{eq:rho_vs_omega_FN}
\end{align}
where $D_s$ is referred to as the spectral dimension
\cite{Nakayama_Yakubo_Orbach1994}.

DNM is, in general, obtained experimentally by spectroscopies
and numerically by normal mode analysis (NMA).
To be relevant to experiments,
we conduct NMA in the all atom model, not in a coarse grained model.
Then, by focusing on the frequency region
corresponding to the residue-residue interaction,
we consider the spectral dimension of the residue network.
We do so because for NMA it is necessary
to take the interaction strengths precisely into account.
In the all atom model, the interaction strengths are quite reliable,
since it is basically obtained from quantum chemical calculations.
In a coarse grained model, in contrast,
the interaction strengths are introduced rather arbitrary.
It is true that the coarse grained models well reproduces the overall
fluctuation of the protein native structure~\cite{Atilgan_etal2001}.
This is, however, largely due to the fact
that only a limited number of lowest frequency normal modes
(or largest amplitude principal components) dominate the fluctuation.
There is no guarantee that they also reproduce DNM for decades.
Indeed, it has been reported that there is an essential difference
in DNM between the all atom model and the coarse grained model
with identical interaction strengths~\cite{Takano_etal2004}.
Instead, here we coarse grain DNM itself,
by truncating the higher frequency region.
We perform NMA by using the program NMODE
implemented in the AMBER software~\cite{Case_etal2005},
with AMBER force field (perm99) and implicit water (Generalized Born) model.
Before NMA, energy minimization is executed
with Newton-Raphson and conjugate gradient method,
so that the norm of the force is less than the order of
$10^{-12}\;\mathrm{kcal}\;\mathrm{mol}^{-1}\mbox{\AA}^{-1}$.

\begin{figure}[tb]
\begin{center}
\includegraphics[width=0.48\textwidth]{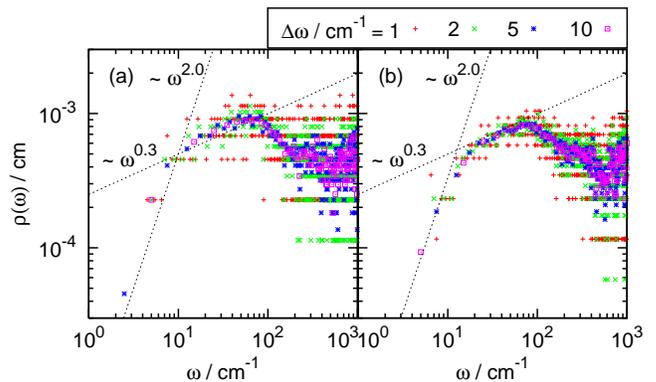}
\caption{Density of normal modes $\rho(\omega)$ of
(a) ribonuclease T1 (PDB ID=9RNT)
and (b) cutinase (1CUS).
Various bin sizes $\Delta\omega$ are taken
so as to display the master curve clearer.}
\label{fig:DNM}
\end{center}
\end{figure}

We have obtained DNM for several proteins, and
\reffig{fig:DNM}
shows typical results;
these are essentially similar to
one of the previous numerical studies~\cite{Yu_Leitner2003}.
There exist two shoulders at around 
$10$ and $100~\mbox{cm}^{-1}$,
which are denoted respectively by $\omega_{FS}$ and $\omega_{GL}$.
The frequency higher than $\omega_{GL}$ corresponds to local motions,
due to covalent-bond stretching and angle bending motions.
The frequency lower than $\omega_{GL}$, in contrast,
corresponds to global motions due to residue-residue interactions,
which we are now interested in.
In the latter region, DNM obeys the power-law scaling
\refeq{eq:rho_vs_omega_FN} with $D_s\approx 1.3$.
At around $\omega_{FS}$, the dimension changes from $1.3$ to $3.0$.
This is due to the finite size effect;
through a long wave-length probe, the protein is regarded just as a 3D object.
Indeed, similar change in slope due to the finite size effect
is observed in percolation clusters~\cite{Nakayama_Yakubo_Orbach1994}.
We expect that, in much larger proteins, $\omega_{FS}$ shift
towards the lower-frequency direction,
and accordingly the region of $D_s\approx 1.3$ becomes wider.
Thus we conclude that residue-residue interaction in proteins
universally follows the power-law scaling
\refeq{eq:rho_vs_omega_FN} with the spectral dimension $D_s\approx 1.3$.

We discuss the reason why some of the previous studies
\cite{Wako1989,benAvraham1993} gave $D_s$ larger than 1.3.
In these studies, $D_s$ was obtained
not from DNM, i.e., the probability density function $\rho(\omega)$,
but from its cumulative distribution function
$\Omega(\omega)=\int_0^\omega \mathrm{d}\omega'\rho(\omega')$.
$D_s$ obtained from $\rho(\omega)$ is identical
with that from $\Omega(\omega)$
if a single scaling holds over the whole range considered.
In proteins, however, the scaling changes at around $\omega_{FS}$
due to the finite size effect.
This accordingly gives an illusionary larger value of $D_s$.
To illustrate this simply, we model the probability density function
as a function that sharply change the scaling at $\omega_{FS}$:
\begin{align}
\rho(\omega)=
\begin{cases}
\displaystyle
\frac{C}{\omega_{FS}}\left(\frac{\omega}{\omega_{FS}}\right)^{D-1}
& (\omega\leq\omega_{FS})\\
\displaystyle
\frac{C}{\omega_{FS}}\left(\frac{\omega}{\omega_{FS}}\right)^{D_s-1}
& (\omega>\omega_{FS})
\end{cases}
\label{eq:prob_exp}
\end{align}
with a dimensionless positive constant $C$.
Its cumulative distribution function is,
\begin{align}
\Omega(\omega) &= \int_0^\omega \mathrm{d}\omega'\rho(\omega') \nonumber \\
&=
\begin{cases}
\displaystyle
\frac{C}{D}\left(\frac{\omega}{\omega_{FS}}\right)^D & (\omega\leq\omega_{FS})\\
\displaystyle
\frac{C}{D_s}\left[\left(\frac{\omega}{\omega_{FS}}\right)^{D_s} - \left(1-\frac{D_s}{D}\right)\right]
& (\omega>\omega_{FS}).
\end{cases}
\label{eq:dist_exp}
\end{align}
The gradient of $\log\Omega$ to $\log\omega$
gives a larger value than the correct spectral dimension $D_s$
at around $\omega\gtrsim\omega_{FS}$.
The gradient would yield $D_s$ in the region
$\omega/\omega_{FS} \gg (1-D_s/D)^{1/D_s}$.
In proteins, $D=3$ and $D_s=1.3$, then $\omega/\omega_{FS} \gg 0.56$.
This region, however, corresponds to the local motions,
not to the global residue-residue interactions
in which we have discovered the universality.

\paragraph{Conclusion: universality class of 3D critical percolation cluster}
We have thus obtained the characteristic dimensions of FN
inherent in the protein native structures,
$(D, D_c, D_f, D_s)=(3, 1.9, 2.5, 1.3)$.
Note that these dimensions are in surprisingly good coincidence
with those in the 3D critical percolation cluster,
$(D, D_c, D_f, D_s)=(3, 1.885, 2.53, 1.3)$~\cite{Stauffer_Aharony1994}.
Hence we here propose that the protein native structures
belong to the universality class of 3D critical percolation cluster.
This is the main statement of this letter.

Then why proteins as residue-contact networks are critically percolated?
Although it is difficult to give the complete answer in the present
stage of this study, still we can provide a purposive explanation
by pointing out two important aspects of proteins; stability and instability.
On the one hand, proteins fold into their own (almost) unique
native structures. Even when they are forced to unfold,
they refold back into the native structures spontaneously
(often with help from molecular chaperons).
In this sense, proteins are stable.
On the other hand, proteins flexibly change their structures.
The structural change is sometimes accompanied with even (partial) unfolding.
In this sense, proteins are unstable.
The coexistence of these two conflicting aspects
is essential for the functions of proteins,
in particular to work as molecular machines or allosteric enzymes.
Being in the critical state is sufficient for that.
Furthermore, the criticality can be even necessary;
proteins should evolve towards the critical state~\cite{Kauffman1993,Bak1996}.
This hypothesis should be verified through the study on molecular evolution,
which is a challenging subject in the future.

\begin{acknowledgments}
This work was partially supported by
Grants-in-Aids for Scientific Research in Priority Areas,
the 21st Century COE Program (Physics of Self-Organization Systems),
and ``Academic Frontier'' Project from MEXT.
\end{acknowledgments}

\end{document}